\documentclass[5p,times]{elsarticle}\journal{Computational Materials Science}\usepackage{hyperref}

\usepackage{amsmath,amssymb}
\usepackage{bm}
\usepackage{graphicx}
\usepackage{caption}
\usepackage{subcaption}
\usepackage{arydshln}
\usepackage[normalem]{ulem}
\usepackage{color}

\definecolor{epcol}{rgb}{0,0, 0.7}
\definecolor{ascol}{rgb}{0 ,0.4, 0}
\definecolor{alertcol}{rgb}{0.7,0, 0}

\newcommand{\mA}{{\mathsf A}}
\newcommand{\mB}{{\mathsf B}}
\newcommand{\mC}{{\mathsf C}}

\newcommand{\calN}{{\mathcal N}}

\newcommand{\bbN}{{\mathbb N}}
\newcommand{\bbR}{{\mathbb R}}

\newcommand{\cut}{{\rm cut}}

\newcommand{\qm}{{\rm qm}}

\renewcommand{\th}{{\rm th}}

\newcommand{\br}{{\bm r}}

\newcommand{\transpose}{{\!\top}}

\newcommand{\QS}[1]{{QS$_#1$}}

\begin{document}

\begin{frontmatter}

\title{Active Learning of Linearly Parametrized Interatomic Potentials}

\author{Evgeny V. Podryabinkin\fnref{myfootnote1}}
\author{Alexander V. Shapeev\fnref{myfootnote2}}
\address{Skolkovo Institute of Science and Technology, Moscow, Russia}
\fntext[myfootnote1]{e-mail: e.podryabinkin@skoltech.ru}
\fntext[myfootnote2]{e-mail: a.shapeev@skoltech.ru}

\begin{abstract}
This paper introduces an active learning approach to the fitting of machine learning interatomic potentials.
Our approach is based on the D-optimality criterion for selecting atomic configurations on which the potential is fitted.
It is shown that the proposed active learning approach is highly efficient in training potentials on the fly, ensuring that no extrapolation is attempted and leading to a completely reliable atomistic simulation without any significant decrease in accuracy.
We apply our approach to molecular dynamics and structure relaxation, and we argue that it can be applied, in principle, to any other type of atomistic simulation.
The software, test cases, and examples of usage are published at \url{http://gitlab.skoltech.ru/shapeev/mlip/}.

\end{abstract}


\end{frontmatter}

\section{Introduction}

Many research areas in materials science, molecular physics, chemistry, and biology involve atomistic modeling.
For example, in molecular dynamics (MD), as a rule, one of the following two classes of interatomic interaction models is used.
The first class is the empirical interatomic potentials---they are very computationally efficient and allow for simulating large atomistic systems for microseconds of simulation time. 
However, they typically yield only qualitative accuracy. 
The other class is quantum-mechanical (QM) models, such as the density functional theory (DFT). 
They are very accurate, but computationally expensive. Their applicability is typically limited to hundreds of atoms and hundreds of picoseconds of simulation time.

Several directions of developing the models that would be both accurate and computationally efficient have been pursued. 
They include the so-called linear scaling DFT \cite{Artacho1999linearDFT,Skylaris2005linearDFT,Bowler2010linearDFT} 
that ensures that the algorithmic complexity grows linearly when the size of the atomistic system increases beyond hundreds of atoms.
Another direction is the development of semi-empirical models, such as the tight-binding model \cite{Finnis2003interatomic}, whose accuracy and efficiency is between those of the empirical potentials and DFT. 
In this paper we pursue a more recent approach based on machine learning. 

\subsection*{Machine learning interatomic potentials} 
Application of machine learning (ML) has recently been put forward as a promising idea that would combine the accuracy of the QM models and the efficiency of the interatomic potentials \cite{ArtrithKolpak2015NNP,GAP2013water,Bartok2010GAP,Behler2011NNP,0953-8984-26-18-183001,BehlerParrinello2007NN,Boes2016NNP-ReaxFF-comparison,Dolgirev2016,Gastegger2015high,Manzhos2015neural,NatarajanMorawietzBehler2015water,Shapeev2016-MTP,GAP2014,Thompson2015316}.
Such machine-learning interatomic potentials (MLIPs) postulate a partitioning of the interatomic interaction energy into individual contributions of the atoms (and sometimes bonds, bond angles, etc.)\ and assume a very flexible functional form for such a contribution, making it a function of the positions of the neighboring atoms, typically with hundreds or more parameters. These parameters are found by requiring the energy, forces and/or stresses predicted by a MLIP to be close to those obtained by a QM model on some atomic configurations. These configurations are called the \emph{training set}, and finding the parameters of a MLIP is known as \emph{training} or \emph{fitting}.
One of the important features of MLIPs are their ability to approximate potential energy surfaces with arbitrary accuracy (at least theoretically) by increasing the number of parameters and the training set.
It should be noted that there are other, ML-based atomistic models of solids, including those predicting the energy directly without partitioning it \cite{Faber2015Coulomb-matrix-crystal,Rupp2012atomization-energies}, or constructing a density functional in a DFT with machine learning \cite{Snyder2012density-functionals}.
A recent overview of ML-based models of materials can be found in \cite{MuellerKusneRamprasad2015ML-review}.

Each of the existing MLIPs has a nontrivial functional form accounting for the physical symmetries of interatomic interaction. 
Namely, a MLIP should be invariant with respect to translation, rotation, and reflection of the space, and also permutation of chemically equivalent atoms. 
In addition, the potential should have a local support (i.e., depend on surrounding atoms only within a finite cut-off radius) and be smooth with respect to atoms coming and leaving the support.
In many instances, it is achieved by designing a fixed number of descriptors \cite{BartokKondorCsanyi2013descriptors,Behler2011symmetry_functions}---scalar functions that satisfy all the symmetries and uniquely encode each atomic environment, and assuming that a MLIP is an arbitrary function (which we call the \emph{regression model}) of these descriptors.
This idea was first put forward by Behler and Parrinello \cite{BehlerParrinello2007NN} proposing an ML model which they called a neural network potential (NNP), based on their descriptors and neural networks as the regression model.
Since then, there has been many works on NNPs, see the review papers \cite{Behler2011NNP,0953-8984-26-18-183001} and references therein, and also more recent works \cite{ArtrithKolpak2015NNP,Dolgirev2016,Gastegger2015high,Manzhos2015neural,NatarajanMorawietzBehler2015water,SmithIsayevRoitverg2017ani,Artrith2016implementation}.
Another group of authors adopted the Gaussian process regression framework \cite{Bartok2010GAP}.
They used the coefficients of spherical harmonics expansion of the smeared atomic positions as descriptors and used the kernel-based ML model, where the kernel was based on the distance between the vectors comprised of those coefficients.
In a follow-up paper, \cite{GAP2014}, the authors refined this idea by proposing the smooth overlap of atomic positions kernel, bypassing the step of designing the descriptors.
For other examples of using Gaussian process regression for constructing interatomic potentials refer to \cite{GAP2013water,DeringerCsanyi2016carbon}.
Three closely related works, \cite{BotuRamprasad2015-MLIP,LiKermodeDevita2015lotf,Glielmo2016force-learning}, use Gaussian process regression to predict the forces on atoms directly, without predicting the energy and taking its gradient.
Finally, \cite{Thompson2015316} proposes a linear regression model with spherical harmonics coefficients as the basis functions.
In the present work, we use the moment tensor potentials (MTPs) \cite{Shapeev2016-MTP}.
These potentials adopt a linear regression model with polynomial-like functions of atomic coordinates as the basis functions.
The MTPs can be interpreted as having descriptors which are based on tensors of inertia of atomic environments.

The MLIPs described above allow for improving their accuracy through increasing the number of the fitting parameters. However, the approximation properties of ML potentials depend not only on their algebraic form, but also on the training set used to fit them.
Choosing a good training set for a potential with many parameters (say, more than ten) proves to be a highly nontrivial practical problem. Indeed, all the existing MLIPs are interpolative, they fail to give reasonable answers outside their training domain. Therefore, a good training set should make a MLIP to be interpolative over all the relevant configurations. Obviously, the more parameters a MLIP involves, the larger and more diverse the training set is required in order to fit such a MLIP.

The problem of choosing a proper training set for the fitting of a reliable MLIP is related to the problem of transferability---the ability of interatomic potentials to extrapolate, i.e., give reasonable predictions outside the training domain (e.g., predict the double vacancy formation energy if only single vacancies are present in the training set).
It is hardly expected that a MLIP can extrapolate beyond the training domain, but even developing a reliable problem-specific MLIP that would accurately interpolate within the training domain is nontrivial, as pointed out, for instance, by Behler \cite[Section 4]{0953-8984-26-18-183001}.
As an illustration of this, the authors of \cite{GAP2014} sampled gamma surfaces (by shifting, in different ways, a part of a crystal along a glide plane) and included them in the training set, which allowed them to compute the properties of dislocations accurately with the exception of their Peierls barrier.
To accurately reproduce the latter they devised a more complicated scheme of generating configurations from the MD trajectories using one version of their potential in order to fit a better version of their potential.

An attractive idea is to attempt to sample the entire space of atomic environments within, for example, a constraint on the minimal interatomic distance.
It is, however, not clear how to do this with sufficient accuracy due to extremely high dimensionality of the space of atomic neighborhoods.
Therefore, in practice, the training set is usually generated by specially designed sampling procedures such as, for example, random perturbations of ideal crystalline configurations \cite{Thompson2015316}, sampling from an ab-initio MD, or a classical MD with empirical potential or another (already fitted) MLIP \cite{GAP2014}.
These sampling procedures, however, do not ensure that the training set covers fully, without ``gaps'', the region in the configuration space required for training MLIPs reliably.
In other words, a potential resulted from such a training procedure may later encounter configurations on which this potential will have to extrapolate.

\subsection*{Active learning and learning on the fly} 
The problem of extrapolation could be resolved if a MLIP were able to detect extrapolative configurations, obtain the QM data for those configurations, and be re-trained.
In this scenario, the extrapolation problem (or the transferability problem) would be solved by reliably predicting on the fly whether a potential is extrapolating on a given configuration.
Alternatively, in the case when learning on the fly cannot be done, the selection of extrapolative configurations can be done offline yielding the training set that improves the transferability of the fitted potential.

Both scenarios are related to a set of ML techniques called active learning (AL). 
In contrast to passive learning in which a potential learns every configuration in the training set, in AL a potential is trained only on a set of selected configurations. 
The key component of any AL method is, thus, its \emph{query strategy}---an algorithmic criterion for deciding whether a given configuration can be treated reliably by an ML model, or we need to re-train our model by querying the QM data for this configuration.
If such decision can be made reliably then, as we show in this paper, we do not have to ensure that the training set generated offline has all the representative configurations.

A general overview of AL approaches can be found in \cite{settles.tr09}.
In the context of interatomic potentials, the first work that proposed AL was \cite{Frederiksen2004Bayesian-active-learning} putting forward a Bayesian query-by-committee strategy. AL was applied by Behler to the neural network potentials \cite[Section 4]{0953-8984-26-18-183001}, using the query by committee-type AL strategy.
Finally, the authors of \cite{BotuRamprasad2015adaptive,Botu2016machine} train a machine learning model predicting the force errors based on the distance between a given atomic configuration and the training set.
A very natural AL approach applicable to force fields based on Gaussian process regression \cite{Bartok2010GAP,GAP2014,GAP2013water,DeringerCsanyi2016carbon,LiKermodeDevita2015lotf,Glielmo2016force-learning}, which has not yet been implemented in practice, would be to use the Bayesian predictive variance, shown to correlate with the actual error, e.g., in \cite{Snyder2012density-functionals}.

In this paper we propose another AL approach for MLIPs
based on the D-optimality criterion 
\cite[Section 3.5]{settles.tr09} 
allowing for detecting the configurations on which a MLIP extrapolates.
This criterion was chosen because there exists an efficient algorithm for checking for D-optimality \cite{oseledets2010how-to1770566}.
Also, as will be discussed in this paper, D-optimality has appealing mathematical interpretations, such as decreasing the uncertainty in determining the parameters or maximizing the volume spanned by the training set in the space of configurations, thus avoiding extrapolation.
We apply our AL approach to the fitting of MTPs, however, it is easily generalizable to a any other linear potential, i.e., a potential whose energy depends linearly on the parameters, such as SNAP \cite{Thompson2015316} or GAP.
In principle, we can apply AL to atomistic systems with any number of chemically different types of atoms, however, most linearly parametrized potentials developed to date are only applicable to systems with a single type of atoms.
We demonstrate that our AL approach allows one to train potentials on the fly with a limited number of QM calculations (occurring, typically, in the initial stage of MD or another atomistic simulation) without loss in accuracy.
In addition, we show that even without learning on the fly, AL can ``optimize'' the training set, in the sense of extracting a significantly smaller subset, training on which reduces the maximal error and improves transferability.

It should be emphasized that the idea of fitting interatomic potentials on-the-fly is not new.
The motivation behind this idea is the same as for the MLIPs---to eliminate expensive QM calculations for those configurations (or atomic neighborhoods) which are similar to the configurations already computed.
Earlier works \cite{Devita1997LOTF,Csanyi2004lotf} proposed a ``learning-and-forgetting'' scheme, in which the interatomic potentials are fitted to the current QM data, and the old QM data are discarded.
A significant step forward was recently done by Li, Kermode and De Vita 	\cite{LiKermodeDevita2015lotf}, who proposed a ``learning-and-remembering'' scheme, in which the database of QM calculations continuously grows with time.
It was demonstrated that this approach allows one to reduce the number of QM calculations by a factor of 30 \cite{LiKermodeDevita2015lotf}.
In all cases, the decision to compute the QM data for a given configuration was taken every $n$ steps (e.g., $n=30$ steps), where $n$ is a fixed number depending on the system, the temperature at which it is simulated, etc.
This is the main difference from the approach proposed in the present work: we formulate a query strategy that is based on geometrical information (atomic positions and supercell vectors) of a configuration and does not use the QM data, thus a well-trained potential will trigger the QM calculations very rarely.

\section{Machine Learning Interatomic Potentials}\label{sec:MLIP}
Let $x$ be a periodic atomistic \emph{configuration} with $N$ atoms in a supercell $L$.
Suppose we can compute, for a given configuration $x$, its QM energy $E^\qm(x)$, forces $f_i^\qm(x)$ ($1\leq i\leq N$) and stresses $\sigma^\qm(x)$.
Such computation typically involves resolving the electronic structure and is very expensive.
For the purpose of our paper, we treat such computation as a black box.

\subsection{Linearly Parametrized Potentials}

We next assume that each atom interacts with its neighborhood defined by a cut-off distance $R_\cut>0$. The neighborhood $\br_i = (r_{i,1},\ldots,r_{i,n})$ of atom $i$ is defined as a collection of vectors pointing from atom $i$ to all the atoms (and their periodic extensions), excluding the atom itself, that are not farther than $R_\cut$.
The number of atoms in the neighborhood, $n$, may depend on $i$.

We assume that the total interaction energy of a configuration is 
\begin{equation} 
\label{eq:energy}
E(x) = \sum_{i=1}^{N} V(\br_i),
\end{equation}
where $V$ is the interatomic potential---a scalar function of the neighborhood $\br_i$.
We define a \emph{linearly parametrized local potential} as having linear dependence on the fitting parameters $\theta_j$: 
\begin{equation}\label{eq:V-def}
V(\br_i) = \sum_{j=1}^m \theta_j B_j(\br_i),
\end{equation}
where $B_j$ are the fixed basis functions. 
The concrete form of the basis functions is not important for what follows, therefore we give the details in the Appendix A on how $B_j$ are constructed for the moment tensor potentials (MTPs) \cite{Shapeev2016-MTP} used in this work.
We define the configuration-dependent basis functions $b_j(x) := \sum_{i=1}^{N} B_j(\br_i)$ and, using \eqref{eq:energy} and \eqref{eq:V-def}, write
\[
E(x) = \sum_{j=1}^{m} \theta_j b_j(x).
\]
The force $f_j(x)$ is a derivative of $E(x)$ with respect to the position of the $j$-th atom, $x_j$:
\begin{equation} \label{eq:force}
f_j(x) = -\nabla_{x_j} E(x), \qquad {1\leq j\leq N},
\end{equation}
and the virial stresses are derivatives with respect to lattice vectors
\begin{equation} \label{eq:stress}
\sigma(x) = \frac{1}{|\det(L)|} \big(\nabla_L E(x)\big)  L^\transpose.
\end{equation}

\subsection{Fitting (training)}

A linearly parametrized potential is uniquely determined by the algebraic form, and the values of the fitting parameters $\theta_j$.
The latter are found through the fitting to the quantum-mechanical energy, forces, and stresses on a set of configurations $X_{\rm TS} = \{x^{(1)},\ldots,x^{(K)}\}$ which we call the \emph{training set}.

The simplest form of fitting is requiring $E\big(x^{(k)}\big) = E^\qm\big(x^{(k)}\big)$.
Expanding the left-hand side yields system of linear algebraic equations on the coefficients $\theta_j$:
\begin{equation} \label{eq:energy_fit}
\sum_{j=1}^m \theta_j b_j\big(x^{(k)}\big) = E^\qm\big(x^{(k)}\big).
\end{equation}
In the matrix notation, we can write this system as $\mA \theta = R$, where
\[
\mA = \begin{pmatrix}
b_1\big(x^{(1)}\big) & \ldots & b_m\big(x^{(1)}\big) \\
\vdots & \ddots & \vdots \\
b_1\big(x^{(K)}\big) & \ldots & b_m\big(x^{(K)}\big) \\
\end{pmatrix}
.
\]
This system is typically overdetermined ($K \geq m$), therefore we define the solution through the pseudoinverse by $\theta := (\mA^\transpose \mA)^{-1} \mA^\transpose R$.

If the configurations in the training set $X_{\rm TS}$ are also provided with the forces $f^\qm(x^{(k)})$ and stresses $\sigma^\qm(x^{(k)})$ one can add two more families of equations in addition to fitting to the energy:
\begin{subequations} 
	\label{eq:f-s_fit}
	\begin{eqnarray}
	C_{\rm f} f_i\big(x^{(k)}\big) = C_{\rm f} f_i^\qm\big(x^{(k)}\big), ~ i = 1, ..., N^{(k)}
	\end{eqnarray}
	\begin{eqnarray}
	C_{\rm s} \sigma\big(x^{(k)}\big) = C_{\rm s} \sigma^\qm\big(x^{(k)}\big).
	\end{eqnarray}
\end{subequations}
These equations are combined into a single least-square minimization problem, therefore we need to introduce the coefficients $C_{\rm f}$ and $C_{\rm s}$ determining the relative importance of the force- and stress-fitting equations relative to the energy-fitting equation.
Expanding the left-hand side of (\eqref{eq:f-s_fit}) yields the following equations:
\begin{subequations} 
	\label{eq:force_stress_fit}
	\begin{eqnarray}
		C_{\rm f}\sum_{j=1}^{m} \theta_j \nabla_{\!x_{i}} b_j\big(x^{(k)}\big) = - 	C_{\rm f} f_i^\qm\big(x^{(k)}\big),	
	\end{eqnarray}
	\begin{eqnarray}
		C_{\rm s} \sum_{j=1}^{m} \frac{\theta_j \big(\nabla_L b_j\big(x^{(k)}\big) \big) L^{(k)\transpose}}{|\det(L^{(k)})|} = 	C_{\rm s}\sigma^\qm\big(x^{(k)}\big),	
		\\
		\qquad i=1,\ldots,N^{(k)}, \quad 1\leq k\leq K.\nonumber
	\end{eqnarray}
\end{subequations}

\section{Active Learning}\label{sec:AL}

Active learning allows one to select the training set from a given set of configurations or a stream of configurations.
It is done only based on the unlabeled data, i.e., no quantum-mechanical energy, forces or stresses are required for making the decision about including a particular configuration $x^*$ into the training set $X_{\rm TS}$.
In the learning-on-the-fly scenario this means that the quantum-mechanical calculations are done only when the configuration is indeed sufficiently ``new''.
The criterion on whether a given configuration $x^*$ should be added to $X_{\rm TS}$ is called the query strategy.

An overview of various query strategies is presented in \cite{settles.tr09}. In the present work we employ the variance reduction query strategy based on the D-optimality criterion \cite[Section 3.5]{settles.tr09} and a fast algorithm, called the maxvol algorithm, developed in \cite{oseledets2010how-to1770566}. 
A way to derive the D-optimality criterion is to assume that the right-hand side of the equations \eqref{eq:energy_fit}, \eqref{eq:force_stress_fit} has a Gaussian random noise and we want to select $m$ out of $K$ equations such that the noise in the solution is minimized.
This is equivalent to choosing a subset of equations \eqref{eq:energy_fit}, \eqref{eq:force_stress_fit} such that its matrix $\mA \in \bbR^{m \times m}$ has the maximal determinant by its absolute value.
Those configurations $x^{(k)}$ that correspond to the selected equations hence form $X_{\rm TS}$.

The active learning algorithm is determined by the following two choices: what equations are used for the fitting and what equations are used for the selection of the configurations.
It should be noted that the two sets of equations need not coincide: for instance, the fitting can be done over all the equations \eqref{eq:energy_fit}, \eqref{eq:force_stress_fit}, but the selection can be done only based on \eqref{eq:energy_fit}.
In the present work we use all the equations for the fitting, and three different versions of the query strategy, as detailed below.

\subsection*{Query strategy \QS{1}: Selection by the energy-fitting equation}

The first query strategy, labeled as \QS{1}, involves only the energy-fitting equation \eqref{eq:energy_fit}.
Given some configuration $x^*$, we need to decide whether to insert it to $X_{\rm TS}=\big\{x^{(1)},\ldots,x^{(m)}\big\}$.
The rows of the matrix $\mA$ corresponding to $x^{(i)} \in X_{\rm TS}$ are:
\[
\mA = \begin{pmatrix}
b_1\big(x^{(1)}\big) & \ldots & b_m\big(x^{(1)}\big) \\
\vdots & \ddots & \vdots \\
b_1\big(x^{(m)}\big) & \ldots & b_m\big(x^{(m)}\big) \\
\end{pmatrix}.
\]
Following the maxvol algorithm \cite{oseledets2010how-to1770566}, we form the row-vector
\begin{equation}\label{eq:row-C}
C :=
\begin{pmatrix}
b_1(x^*) & \ldots & b_m(x^*)
\end{pmatrix}
\mA^{-1}.
\end{equation}
If we replace the $k$-th row of $\mA$ by $\big(b_1(x^*) \ldots b_m(x^*)\big)$ then $|{\rm det}\,\mA|$ will change by a factor of $|C_k|$
(the easiest way to see this is to use the Cramer's formula).

Our query strategy is thus as follows.
If
\begin{equation}\label{eq:theta-criterion}
\gamma(x^*) := 
\max_{1\leq k \leq m} |C_k|> \gamma_\th,
\end{equation}
where we call $\gamma_\th \geq 1$ a \textit{threshold},
then we add $x^*$ to the training set and remove $x^{(k)}$ with $k = {\rm arg\,max}_k |C_k|$; otherwise, we keep the $X_{\rm TS}$ as is.
This procedure guarantees that if $x^{(k)}$ is replaced by $x^*$ in $X_{\rm TS}$ then $|{\rm det\,} \mA|$ is increased at least by a factor of $\gamma_\th$.
If in the algorithm one stores $\mA^{-1}$ instead of $\mA$ then the complexity of computing $\gamma(x^*)$ is only $O(m^2)$.
When $\mA$ changes, it also takes $O(m^2)$ operations to update $\mA^{-1}$ using the Sherman-Morrison \cite{rank-1-update} rank-one update.

It is worthwhile noting that the elements of $C$ can be interpreted as the coefficients of expressing $E(x^*)$ through $E\big(x^{(k)}\big)$:
\[
E(x^*) = \sum_{k=1}^{K} C_k E\big(x^{(k)}\big).
\]
Hence we can say that if all $|C_k|\leq 1$ then we are \emph{interpolating} the predicted value of the energy, $E(x^*)$, through the energies $E\big(x^{(k)}\big)=E^\qm\big(x^{(k)}\big)$.
Hence, $\gamma(x^*)$ defined by \eqref{eq:theta-criterion} has a meaning of a degree of extrapolation that we commit when evaluating $E(x^*)$.
Hence we call $\gamma(x^*)$ the \emph{extrapolation grade} and the parameter $\gamma_\th \geq 1$ defines the maximal allowed extrapolation grade.
It should be emphasized that $\gamma(x^*)$ does not depend on the QM data and is therefore a geometric feature of the configuration $x^*$ and configurations from $X_{\rm TS}$.

\subsection*{Query strategy \QS{2}: Selection by all equations}

The second query strategy, \QS{2}, the matrix $\mA$ is formed by $m$ equations chosen from \eqref{eq:energy_fit}, \eqref{eq:force_stress_fit}.
Thus, each configuration $x^*$ yields $1+3N+6$ rows in place of \eqref{eq:row-C},
\[
\mC =
\underbrace{
	\begin{pmatrix}
	b_1(x^*) & \ldots & b_m(x^*) \\
	C_{\rm f} \nabla_{x_1} b_1(x^*) & \ldots & C_{\rm f} \nabla_{x_1} b_m(x^*) \\
	\vdots & \vdots & \vdots \\
	\end{pmatrix}
}_{=:\mB}
\mA^{-1}
,
\]
(here $\nabla_{x_1} b_1(x^*)$ comes from expanding the force in basis functions)
and it is selected for training if
\begin{equation}
\max_{k,j} |\mC_{kj}|>\gamma_\th.
\end{equation}
We then use the maxvol algorithm \cite{oseledets2010how-to1770566} to maximize $|{\rm det}\,\mA|$; it is a greedy algorithm replacing rows of $\mA$ with rows of $\mB$ until $\max_{k,j} |\mC_{kj}| \leq \gamma_\th$.
The algorithm is detailed in Appendix B.
Note that in this query strategy there may be more than one row of $\mA$ corresponding to one configuration, and thus less than $m$ configurations may be selected into the training set.
The algorithm complexity is $O(N m^2)$.

\subsection*{Query strategy \QS{3}: Selection by neighborhoods}

To formulate the last approach we suppose, for the sake of argument, that we could also fit to the site energies,
\begin{eqnarray}
\label{eq:site_en}
V\big(\br_i^{(k)}\big) \equiv
\sum_{j=1}^m \theta_j B_j\big(\br_i^{(k)}\big) = {V_i^\qm}^{(k)} \nonumber
\\
\qquad 1\leq k\leq K,~~ 1\leq i\leq N^{(k)}
.
\end{eqnarray}
Thus, in the third approach, \QS{3}, $\mA$ is formed by the equations \eqref{eq:site_en}.
We proceed similarly to \QS{2} and compose
\[
\mC =
\underbrace{
	\begin{pmatrix}
	B_1(\br_1^*) & \ldots & B_m(\br_1^*) \\
	\vdots & \vdots & \vdots \\
	B_1(\br_N^*) & \ldots & B_m(\br_N^*) \\
	\end{pmatrix}
}_{=:\mB}
\mA^{-1}
.
\]
Then if $\max_{k,j}|\mC_{kj}|>\gamma_\th$ then we replace the rows in $\mA$ with the rows in $\mB$ and update $X_{\rm TS}$ accordingly.
As in \QS{2}, there may be more than one row of $\mB$ corresponding to one configuration, and thus the training set may contain less than $m$ configurations.	The algorithm complexity is the same as that of \QS{2}, $O(N m^2)$.
It should be emphasized that no site energies are actually required from quantum mechanical data for the implementation of \QS{3}.

\subsection*{Learning on the fly}

\begin{figure}[htbp]
	\centering
	\includegraphics[width=2.5in]{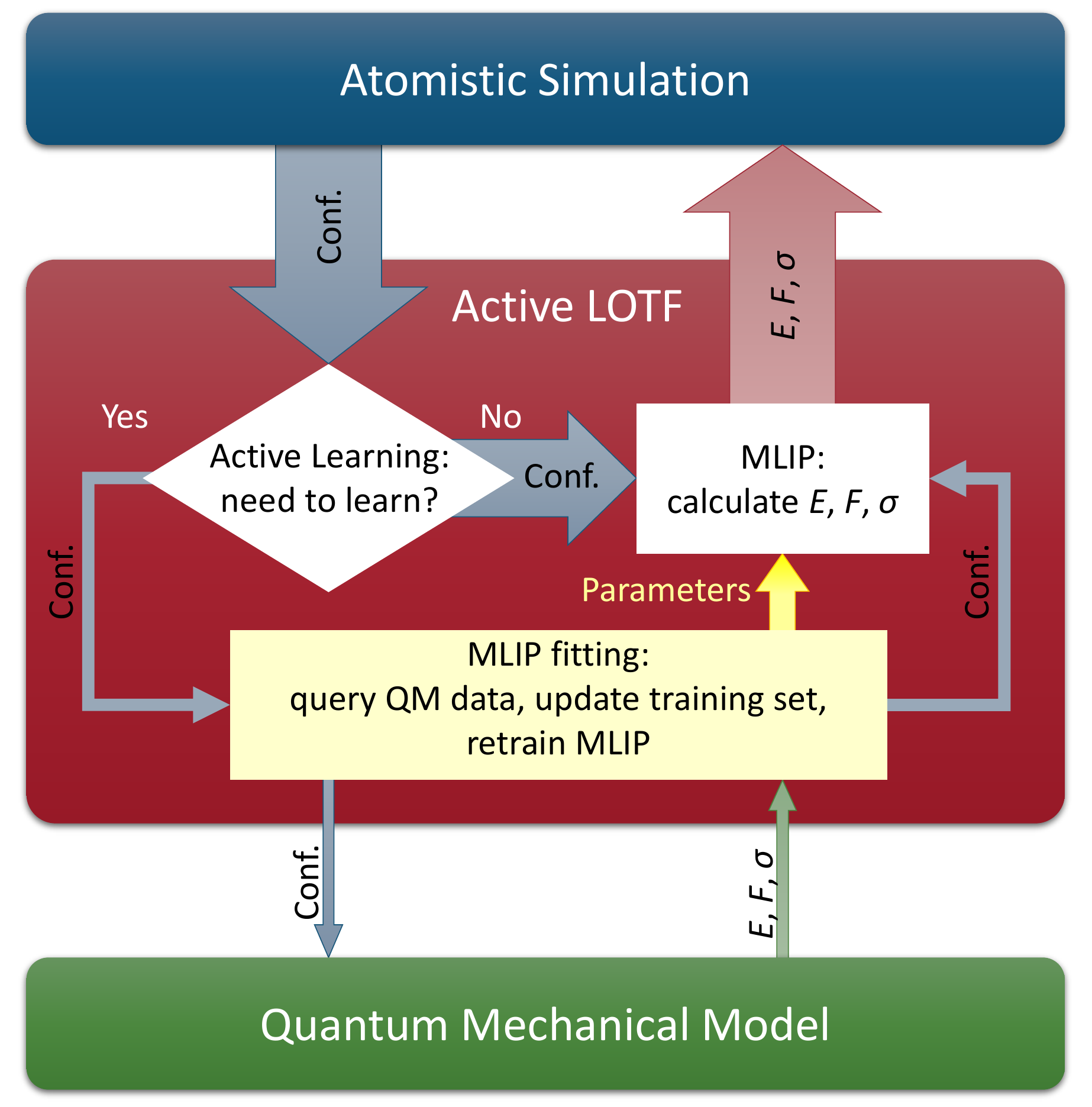}
	\caption{Workflow in actively learning a potential on the fly.
		An active-learning scheme gets an atomistic configuration and returns its energy, forces, and stresses, by possibly retraining the interatomic potential.
	}
	\label{fig:Diagram}
\end{figure}

The AL methodology naturally applies to learning on-the-fly scenario that combines the interatomic potential evaluation and its learning into a single routine, see Fig.~\ref{fig:Diagram}.
At each iteration of an atomistic simulation, an unlabeled configuration $x^*$ comes as an input to an AL procedure, that does the following.
\begin{enumerate}
	\item[1.]  Calculate extrapolation grade,  $\gamma(x^*)$. If $\gamma(x^*) \leq \gamma_\th$ then go to step 5, else
	\subitem{2.} Calculate $E^\qm(x^*)$, $f^\qm(x^*)$, and $\sigma^\qm(x^*)$ with a quantum-mechanical model.
	\subitem{3.}  Update $X_{\rm TS}$ (and hence $\mA$) with $x^*$.
	\subitem{4.}  Re-fit the MLIP and obtain new $\theta_1, \ldots, \theta_m$.
	\item[5.]  Return $E(x^*),f(x^*),\sigma(x^*)$ according to the current values of $\theta_1, \ldots, \theta_m$.
\end{enumerate}

Note that in this scheme the parameter $\gamma_\th$ controls the efficiency of the learning-on-the-fly scheme, effectively ignoring configurations which increase $|{\rm det\,}\mA|$ only slightly and perform expensive QM calculations only for sufficiently ``new'' configurations.
In practice, there is an optimal range of $\gamma_\th$ for which the QM calculations are not done too often, and on the other hand, the extrapolation does not significantly decrease the accuracy of the potential, see Section~\ref{sec:test:lotf}.

As another application, AL can be applied to reducing the training set, for instance, when it contains many similar configurations.
In Appendix C we show that such offline application of AL improves the transferability of a MLIP and reduces maximal errors as compared to learning from the full database.

\section{Testing}\label{sec:testing}

In this section we test the proposed AL schemes.
In Section~\ref{sec:test:1d} we give an illustratory example of how AL works for a system with one degree of freedom.
Then, in Section~\ref{sec:test:accuracy} we test the accuracy of the fitting of the MTP potentials \cite{Shapeev2016-MTP} on a crystalline Li system, and in Section~\ref{sec:test:corr} we show that the extrapolation grade $\gamma$ correlates with the error of fitting.
Finally, in Section~\ref{sec:test:lotf}, we will test learning on the fly.

All the tests are done using the open-source software that we publish at \url{http://gitlab.skoltech.ru/shapeev/mlip/}.
The distribution package includes the examples of applications described in this section.

\subsection{A one-dimensional illustration}\label{sec:test:1d}

\begin{figure}[htbp]
	\centering
	\includegraphics[scale=.8]{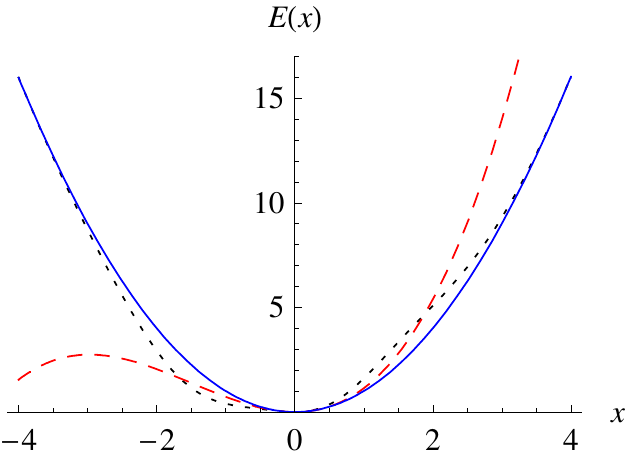}
	\caption{A one-dimensional example. The dotted line is the exact energy $E^\qm(x)$, the red dashed line is the least-mean-square fit, and the blue solid line is the fit with active learning.
		The least-mean-square has the lowest possible energy (i.e., is closer to the exact energy in the energy well), but creates a spurious energy barrier around $x=-3$ that changes the correct long-term behavior of the system.
	}
	\label{fig:1dfits}
\end{figure}

We start by illustrating how the proposed method works in a simple one-dimensional example.
Suppose our system has only one degree of freedom, $x\in[-4,4]$, and is described by the energy $E^\qm(x) = x^2 + x^3 e^{-x^2/2}$, shown in Figure~\ref{fig:1dfits}, dotted line.
We approximate it by a potential $E(x) = \theta_1 x^2 + \theta_2 x^3$.
Suppose that we can sample the exact finite-temperature MD arbitrarily long, and we minimize the mean-square error of the fit on these MD samples.
This is equivalent to minimizing the free-energy functional
\[
\int_{-4}^4 (E(x) - E^\qm(x))^2 e^{-E^\qm(x)} {\rm d}x
\]
resulting from a Boltzmann distribution with the dimensionless temperature $k_{\rm B} T = 1$.
We then obtain the fit $E(x) = 0.93 x^2 + 0.21 x^3$, shown in Figure~\ref{fig:1dfits}, red dashed line.
The root-mean-square error of this fit, defined as the standard deviation of $E(x)-E^\qm(x)$ with respect to the Boltzmann distribution, is only $0.25$. However, the major problem is that this fit creates a finite energy barrier around $x=-3$, which would cause the system to occasionally escape the physical region and drive the system to a spurious minimum $x\to-\infty$.

We next apply the AL approach to this problem by selecting two points (since there are two basis functions) from the interval $[-4,4]$ for the fitting of $E(x)$.
The most optimal points are $x_1 = -4$ and $x_2=4$, as they maximize $\det\begin{pmatrix}
x_1^2 x_1^3 \\ 	x_2^2 x_2^3
\end{pmatrix}.$
The fit is then $E(x) = x^2 + 3\cdot10^{-4} x^3$, shown in Figure~\ref{fig:1dfits}, blue solid line.
Its error is $0.46$, but it correctly predicts the barriers at the boundary of the region of interest, and hence the MD will not escape the region of interest.

It thus can be concluded that, at least in this one-dimensional example, AL offers reliability at the cost of a trade-off in accuracy as compared to passive learning.
In the following subsections we will see that the difference between passive and active sampling is even more pronounced in a realistic MD---AL offers in practice a completely reliable model at the cost of a marginal error increase.

\subsection{Accuracy of learning molecular dynamics}\label{sec:test:accuracy}

In this and the following subsections we perform atomistic simulations of Lithium.
The tests are performed in a cubic supercell of 128 Lithium atoms arranged in a b.c.c.\ lattice.
The length of the supercell in each direction is greater than twice the cut-off radius, $2R_\cut = 10$\AA.
This ensures that each atomic neighborhood does not contain multiple periodic images of a single atom.

The energies, forces, and stresses were computed using DFT with the VASP code \cite{VASP1,VASP3,VASP4}, a projected augmented wave (PAW) pseudopotential \cite{Blochl1994PAW}, and the Perdew-Burke-Ernzerhof exchange-correlation functional \cite{PBE}.
Lithium is an alkaline metal with one valence electron and therefore its electronic structure can be computed faster than for the other elements, which is helpful in collecting large statistics for the tests.

Before testing our AL approach, we perform a test of accuracy of the MTP potential for Li by fitting to a fixed quantum-mechanical database.
The database was comprised of four ab-initio MD trajectories sampling an NVT-ensemble at temperature $T=300$\,K, each trajectory ran for 6\,000 time steps, each time step was 1 fs.
A sequence of MTPs with different number of basis functions, $m$, was generated; the more basis functions are included, the better is the accuracy.
The fitting errors are given in the Table~\ref{tab:table1}. Here and in what follows we report the root-mean-square (RMS) error and the maximum error.
As can be seen, the potentials systematically converge, however, increasing the number of basis functions beyond 100, essentially, does not reduce the error.
Therefore the results for the subsequent tests will be quoted for the same set of 100 basis functions.

%
\begin{table}
	\centering
	\caption{RMS fitting errors in energy, forces and stresses for MTPs with different number of basis functions, $m$. The root-mean-square (RMS) and the maximum errors are quoted.
	}
	\label{tab:table1}
		\begin{tabular}{rlllll}
		\hline
			$m$ & Energy error & \multicolumn{2}{c}{Force error}& \multicolumn{2}{c}{Stress error} \\ 
			& (meV/atom) & (eV/\AA) & (\%) & (GPa) & (\%) \\ 
			\hline
			10      & ~~~~~$0.35$ & 0.023 &  7.4  & 0.073 &7.0\\
			30      & ~~~~~$0.22$ & 0.018 &  5.8  & 0.060 &5.8\\
			100     & ~~~~~$0.19$ & 0.016 &  5.1  & 0.052 &5.2\\
			300     & ~~~~~$0.17$ & 0.015 &  4.9  & 0.045 &4.4\\
			1000    & ~~~~~$0.15$ & 0.015 &  4.8  & 0.040 &3.8\\ 
		\hline
		\end{tabular}
\end{table}

\subsection{Correlation of the error and the extrapolation grade}\label{sec:test:corr}

We next show that the force error correlates with the extrapolation grade $\gamma(x^*)$.
Note that the cost of evaluating $\gamma(x^*)$ is as cheap as a matrix-vector multiplication (or a matrix-matrix multiplication for \QS{2} and \QS{3}), as we do not need to consider the cost of evaluating $\begin{pmatrix}
b_1(x^*) & \ldots & b_m(x^*)
\end{pmatrix}$ in \eqref{eq:row-C} (and $\mB$ for \QS{2} and \QS{3}) since it is required for computing $E(x)$ in any case.
The correlation between the error and $\gamma(x^*)$ may be used to assess the applicability of a MLIP to a given configuration $x^*$ during an atomistic simulation.
Even more important than simply knowing $\gamma$ for configurations appeared in MD, we can store the configurations with high $\gamma$ in order to perform the QM calculations and refit MLIP on them on-the-fly or after the simulation.

\begin{figure}[htbp]
	\centering
	\includegraphics[width=0.95\columnwidth]{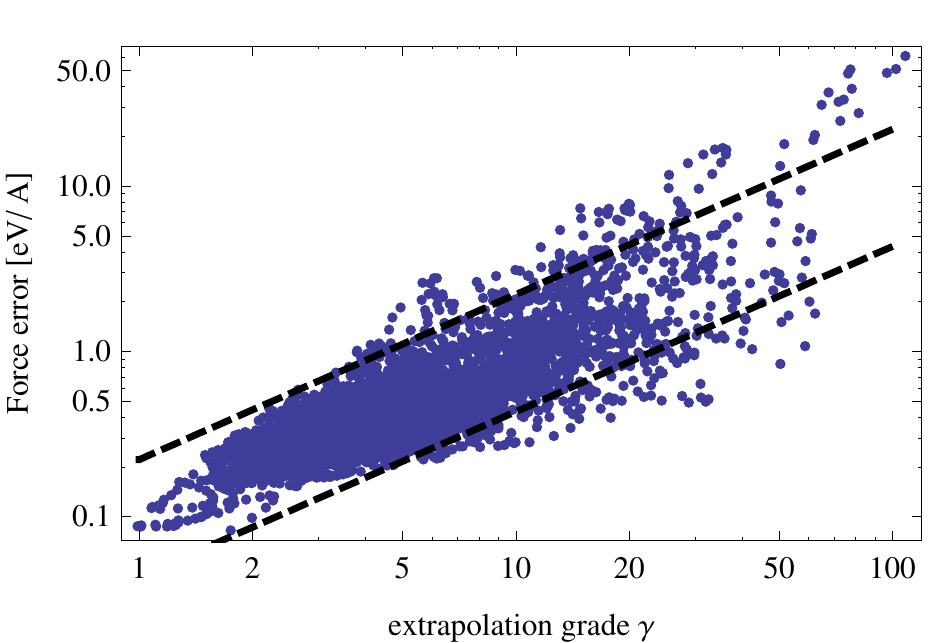}
	\caption{Correlation between the extrapolation grade $\gamma(x)$ and the force error $\Delta f(x) = \big( \frac1N \sum_{i=1}^N |f_i(x) - f_i^\qm(x)|^2 \big)^{1/2}$.
		Each point on the graph corresponds to an MD time step.
		For 95\% of configurations the RMS force error is within $[0.04\,\gamma,0.22\,\gamma] \, {\rm eV/\AA}$ (dashed lines)---this indicates a good correlation between the error and the extrapolation grade.}
	\label{fig:transferability}
\end{figure}

We compute the force errors and the extrapolation grade each time step of an MD at $T=300$\,K.
The force errors versus the extrapolation grade are plotted in Figure~\ref{fig:transferability}.
A good correlation between the two can be observed---this indicates that in practice an extrapolation grade can predict the correct order of magnitude of the error a potential makes on a given configuration without performing a QM calculation.

\subsection{Learning on the Fly}\label{sec:test:lotf}

We next test our AL strategy in a realistic setting of MD and structure relaxation.

\begin{figure*}[ht]
		\hfill
\begin{subfigure}[t]{0.49\textwidth}
	\includegraphics[scale=.23]{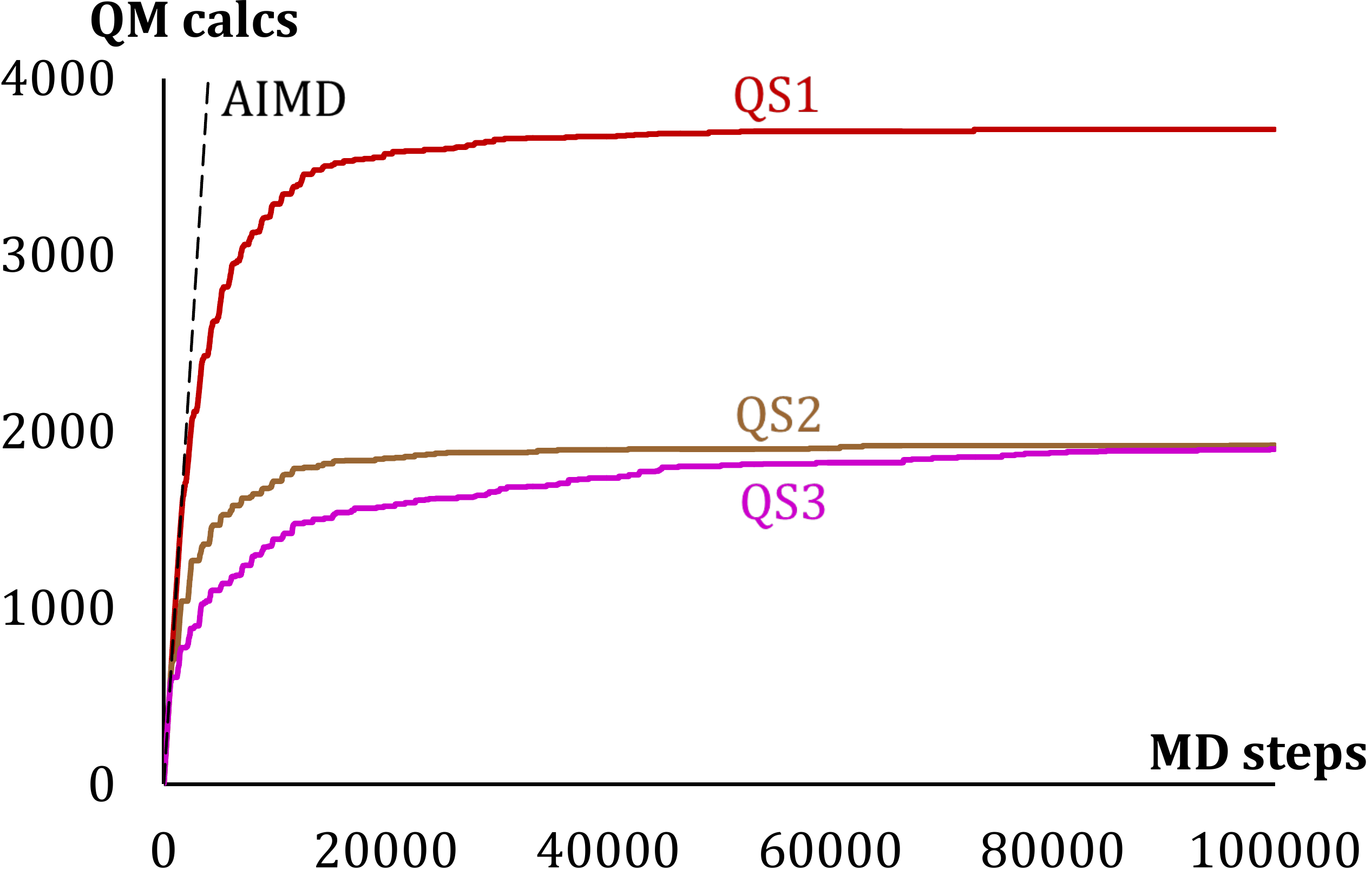}
	\caption{\footnotesize Comparison of the query strategies ($\gamma_\th=1$).
	}
	\label{fig:lotf:1}
\end{subfigure}
		\hfill
\begin{subfigure}[t]{0.49\textwidth}
	\includegraphics[scale=.23]{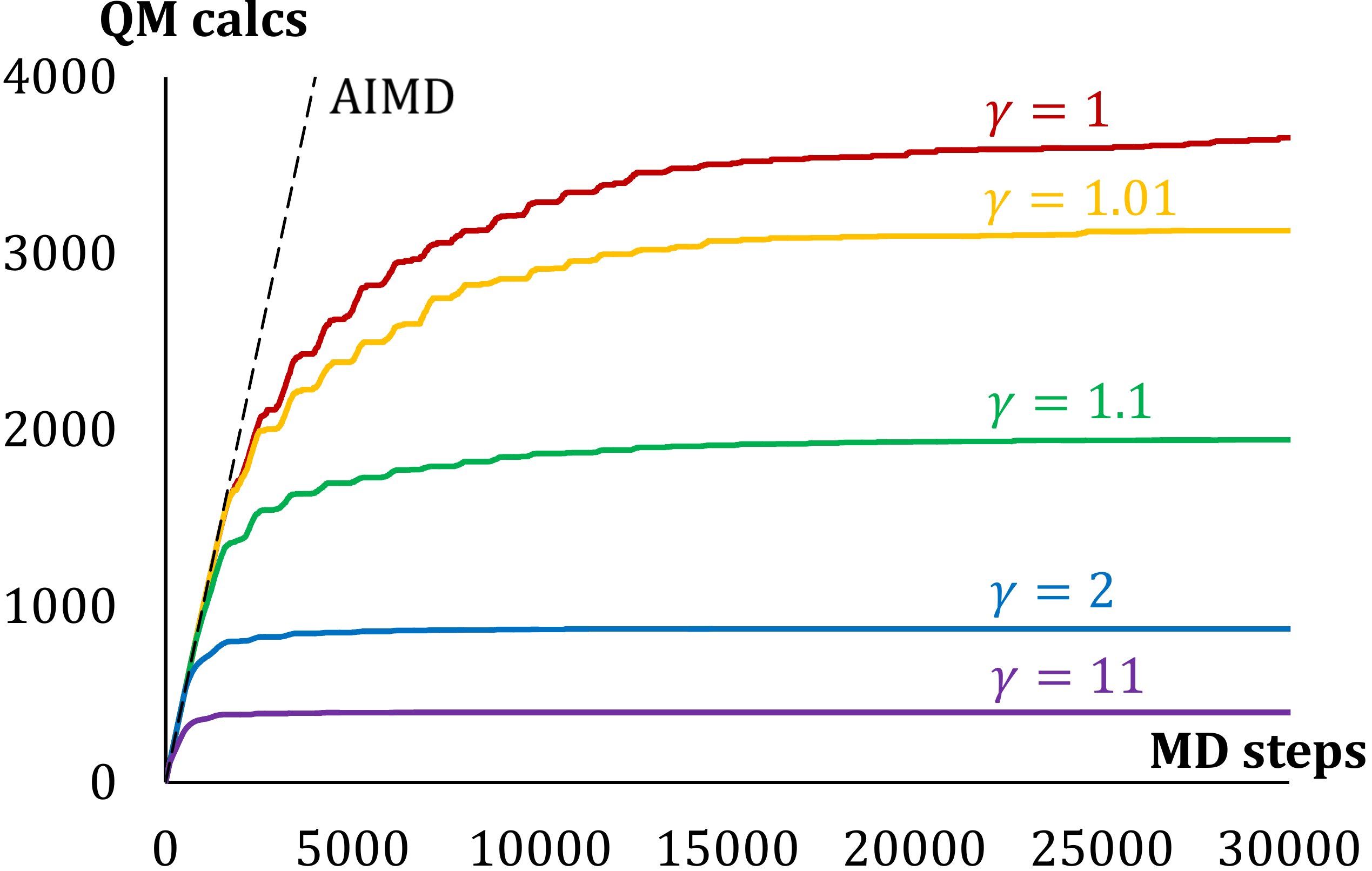}
	\caption{\footnotesize 
		\QS{1} with different thresholds $\gamma_\th$ \\ (first 30 ps).
	}
	\label{fig:lotf:2}
\end{subfigure}
	\caption{Amount of QM calculations in a learning-on-the-fly MD as a function of the MD time step. (a): Comparison of the query strategies; (b): \QS{1} with different thresholds $\gamma_\th$  (first 30 ps).
	}
	\label{fig:lotf}
\end{figure*}

We run MD trajectories at $T=300$\,K for 100 ps, training an MTP on the fly.
Graphs in Figure~\ref{fig:lotf}(a) show the amount of QM calculations as a function of the simulation time. 
One can see that all query strategies require many QM calculations during an initial phase (1--5ps) and then gradually move to the regime when QM calculations are required only rarely. \QS{1} requires about twice more QM calculations as compared to \QS{2} and \QS{3}.

As one can see from the Figure~\ref{fig:lotf}(b) the amount of the QM calculations drops significantly as $\gamma_\th$ increases (with $\gamma_\th=2$ about four times less QM calculations are required as compared to $\gamma_\th=1$).
On the other hand, as seen from Table~\ref{tab:table7}, the errors essentially do not increase up to $\gamma_\th=2$ and only at $\gamma_\th=11$ the maximum error exhibits a slight increase (for $\gamma_\th>1$ only the errors of \QS{1} are tabulated, however, the behavior of \QS{2} and \QS{3} with increasing $\gamma_\th$ is essentially the same).
This indicates that one can easily tune the efficiency-versus-accuracy performance of an AL scheme by adjusting $\gamma_\th$. 
Based on our experience, we find that a value for $\gamma_\th$ between $2$ and $11$ is a good choice in practice---it does not significantly reduce the accuracy, while the number of the QM calculations is just a few times higher than the theoretical minimum (which is equal to the number of undetermined parameters).

\begin{table}[htbp]
	\centering
	\caption{Accuracy for different query strategies and $\gamma_\th$.}
	\label{tab:table7}
		\begin{tabular}{rrc@{~~~}cc}
		\hline
			Query & & $\#$QM & \multicolumn{2}{c}{force error} \\
			strategy &$\gamma_\th$  & calcs & \multicolumn{2}{c}{(eV/\AA)} \\
			& & & RMS & max \\
			\hline
			\QS{1}& 1	&3712	& 0.015  & 0.13\\
			\QS{1}& 1.01&3168	& 0.015  & 0.10\\
			\QS{1}& 1.1	&1958	& 0.016  & 0.19\\
			\QS{1}& 2	&873	& 0.016  & 0.13\\
			\QS{1}& 11	&397	& 0.016  & 0.31\\ 
			\hline
			\QS{2}& 1	&1921	& 0.016  & 0.09\\
			\hline
			\QS{3}& 1	&1900	& 0.015  & 0.12\\
		\hline
		\end{tabular}
\end{table}

\subsubsection*{Comparison with classical learning on the fly}	

Next we test the reliability of our AL strategy in a scenario of learning-on-the-fly MD for bulk Li, as in the previous test case.
We compare it to a the classical learning-on-the-fly algorithm inspired by \cite{Devita1997LOTF,Csanyi2004lotf,LiKermodeDevita2015lotf}.
In \cite{LiKermodeDevita2015lotf} the authors propose to: (1) learn from an initial, one or few picosecond-long AIMD trajectory, and (2) perform an MD with the trained potential, additionally adding configurations to the training set once in every $N_{\rm step}$ time steps.
For the purpose of illustration, we choose $N_{\rm step}=100$ and run MD at the melting temperature, although we note that the authors of \cite{LiKermodeDevita2015lotf} report their results for $N_{\rm step}=30$ at a much lower temperature.

\begin{figure}[htbp]
	\centering
	\includegraphics[scale=.75]{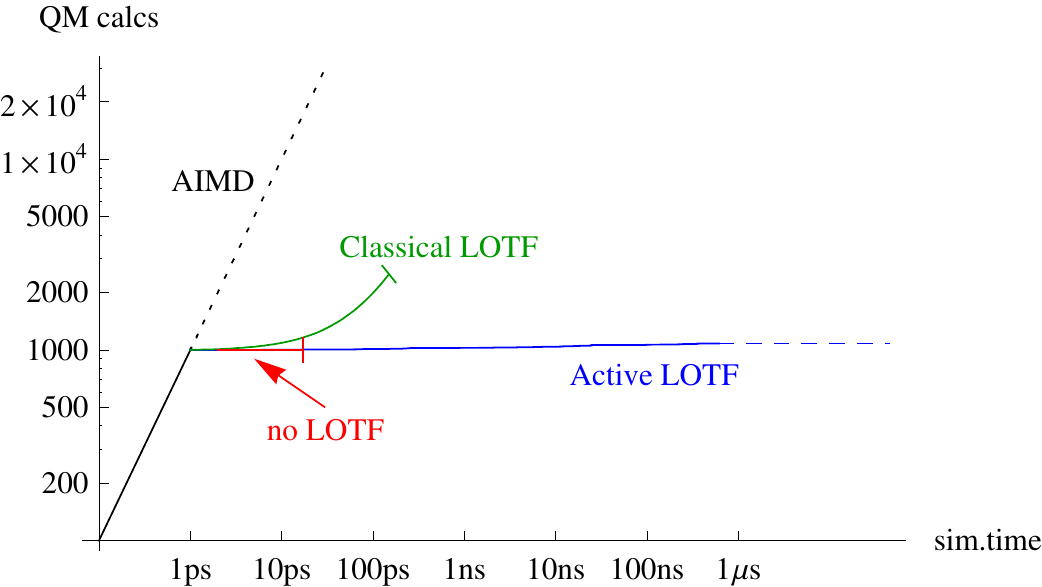}
	\caption{%
		Comparison of {\it ab initio} molecular dynamics (AIMD) with no-learning MD, classical learning on the fly (LOTF) inspired by \cite{LiKermodeDevita2015lotf}, and active LOTF.
		The no-learning and classical LOTF MD are not completely reliable: on average every 15ps the no-learning MD fails, i.e., escapes into an unphysical region in the phase space.
		The classical LOTF makes this ten times more reliable (failure time of 150ps) at the expense of extra 1500 QM calculations.
		In contrast, the active LOTF makes MD completely reliable (i.e., failures are not observed) at the cost of only 50 QM calculations as measured over the first $0.5\mu{\rm s}$ of simulation time.
	}
	\label{fig:learning-comparison}
\end{figure}

The results of this test are illustrated in Figure~\ref{fig:learning-comparison}.
If a potential is trained on a fixed database, it is observed that once in about $15{\rm ps}$ the atomistic system escapes into an unphysical region characterized by very low (below 1\AA) bond lengths.
Therefore, to assess the reliability of a potential, we terminate the MD if after some simulation time the minimal distance between atoms becomes smaller than $1.5{\rm \AA}$.
We call the simulation time after which half the trajectories are terminated (i.e., the trajectory half-life), the \emph{failure time}.
From the transition state theory, we estimate that in an AIMD, the failure time is of the order of $10^{10} {\rm s}$---which is much larger than is accessible even with a classical MD.

Figure~\ref{fig:learning-comparison} illustrates the comparison of classical and active learning.
The classical learning-on-the-fly scheme inspired by \cite{LiKermodeDevita2015lotf}, increases the average failure time from about $15{\rm ps}$ to $150{\rm ps}$ at a cost of 1500 additional QM calculations.
In contrast, with the proposed active learning-on-the-fly scheme, we ran effectively a 0.5$\mu$s-long simulation, which has not failed a single time.
We used parallel replica method \cite{Voter1998parrep} to access such a long timescale.
During this 0.5$\mu$s-long trajectory the scheme required only about 50 additional QM calculations (out of them only 5 during the first $150{\rm ps}$, as compared to 1500 for the classical learning).
Thus, the AL approach offers, in practice, a completely reliable scheme as opposed to the classical learning approach at a much smaller cost.

We note that we have observed many classical learning-on-the-fly MD trajectories that fail within the first 5ps of learning of the fly.
This means that, in the logic of \cite{LiKermodeDevita2015lotf}, the initial MD trajectory should be extended beyond 1ps, at the cost of more QM calculations.
Instead, in this work we simply discarded those trajectories when estimating the failure time (otherwise the estimated failure time would be significantly smaller than 150ps).
It should also be remarked that decreasing $N_{\rm step}$ from 100 to 30, as suggested in \cite{LiKermodeDevita2015lotf}, should further improve the reliability of the classical learning on the fly, but this would further increase its cost and will make its failure time more expensive to measure.

\subsubsection*{Automatic expansion of the training region}

In the next test, we illustrate how the AL scheme allows for an automatic expansion of the region spanned by the training set.
We start with the potential from the previous test, trained on the fly for 500ps at $450K$.
The training set contains 100 crystalline configuration.
We then start a new learning-on-the-fly MD at $900K$ which is well above the melting point.

\begin{figure}[htbp]
	\centering
	\includegraphics[scale=.75]{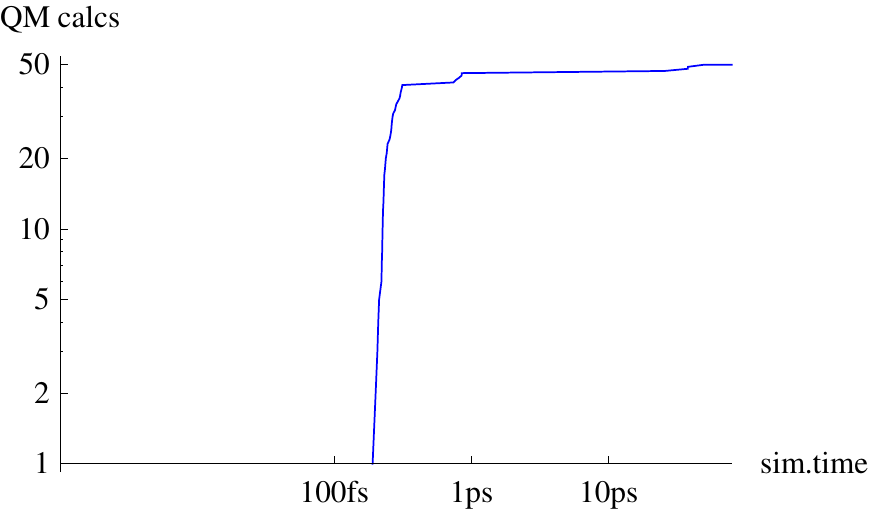}
	\caption{%
		Learning after raising the temperature from 450K to 900K.
		The atomistic system takes about 200fs to liquify, and for the next 100fs the potential does most of learning of liquid configurations.
		After this, the QM calculations are done only occasionally.
	}
	\label{fig:liquid}
\end{figure}

The performance of learning on the fly is shown in Figure~\ref{fig:liquid}.
The solid-to-liquid transition occurs after about 200fs of simulation time and most of learning takes place between 200fs and 300fs.
We emphasize that the AL algorithm does not ``know'' of the temperature change---it makes the decision only based on incoming atomistic configurations.

\begin{table}
	\centering
	\caption{Errors of potentials trained at different temperatures when tested on configurations sampled at different temperatures.
	The absolute errors are in eV/\AA.
	The relative errors are also given in parenthesis.
	The potential trained on crystalline configurations (at $300K$ or $450K$) fails on the liquid configurations ($900K$).
	When additionally trained on liquid configurations, the potential shows somewhat higher errors for crystalline configurations, but no longer fails on liquid configurations.
	}
	\label{tab:3x3-errors}
		\begin{tabular}{r:ccc}
		\hline
		          & \multicolumn{3}{c}{Force error at:} \\
		Potential & $300K$ & $450K$ & $900K$ \\
		\hline
		MTP$_{300}$ & 0.016 (4.6\%) & 0.022 (5.7\%) & 12.1\phantom{0 (0.0\%)} \\
		MTP$_{450}$ & 0.017 (4.7\%) & 0.020 (5.2\%) & 11.7\phantom{0 (0.0\%)} \\
		MTP$_{900}$ & 0.030 (8.4\%) & 0.033 (8.5\%) & 0.062 (7.0\%) \\
		\hline
		\end{tabular}
\end{table}

After learning at $900K$, the prediction errors at lower temperatures somewhat increase.
To test this, we actively selected three sets of configurations, sampled at $300K$, $450K$, and $900K$, and fitted three potentials on these sets, denoted by MTP$_{300}$, MTP$_{450}$, and  MTP$_{900}$, respectively.
The errors of these potentials on these sets are shown in Table \ref{tab:3x3-errors}.
The potentials trained on crystalline configurations (at $300K$ or $450K$) fail to predict forces for liquid configurations.
Nevertheless, after additional training on liquid configurations, the potential became applicable to both, liquid and crystalline configurations; however, the errors on crystalline configurations became somewhat larger.

\subsubsection*{Active learning beyond molecular dynamics}

\begin{figure}[htbp]
	\centering
	\includegraphics[scale=.25]{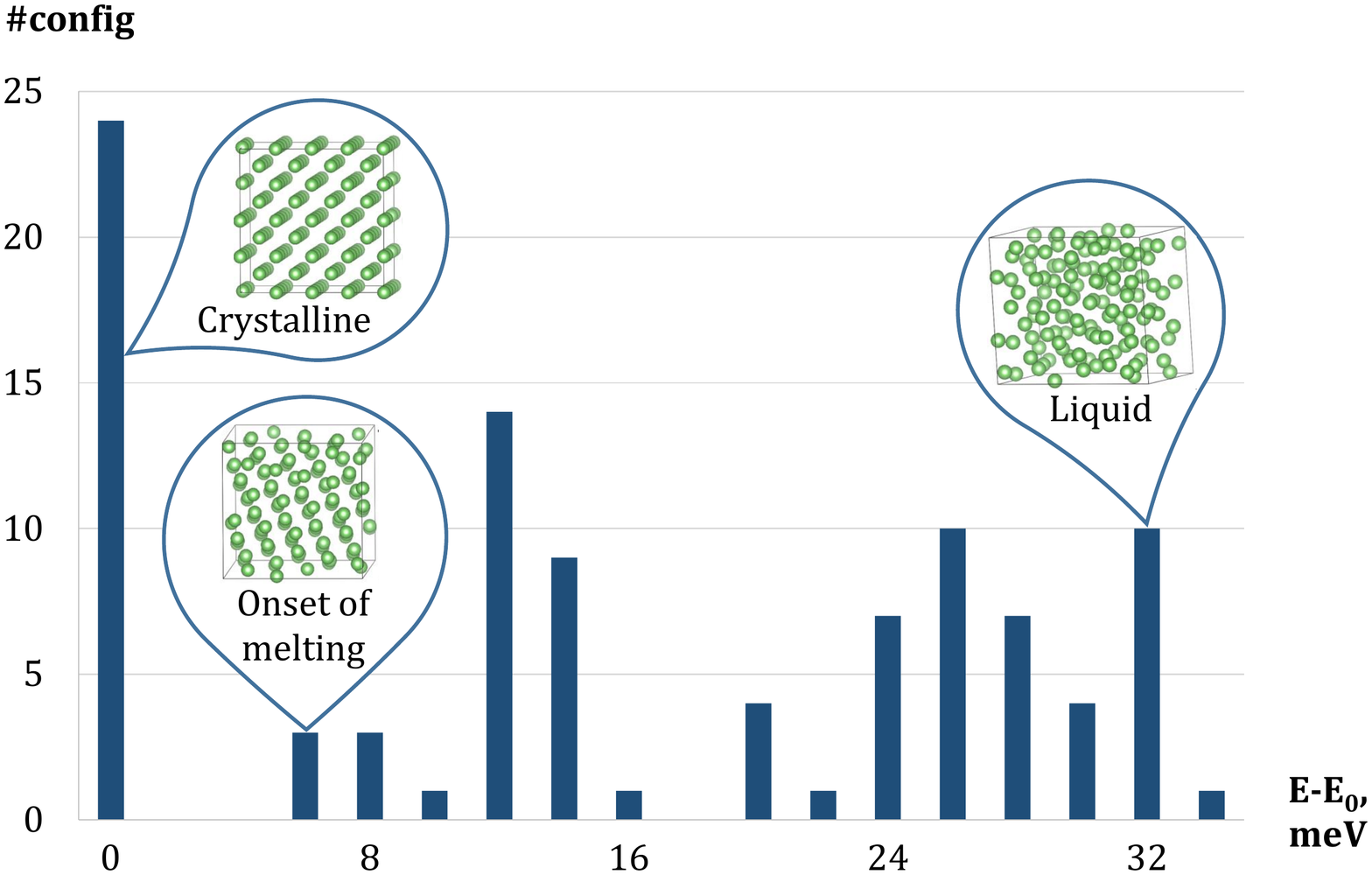}
	\caption{%
		An illustration of configurations selected into the training set by active learning. The X-axis is the range of energies per atom of configurations after relaxation.
		The Y-axis is the number of configurations within a certain energy range.
		The training set features 24 crystalline configurations and 76 fully or partly liquid.
	}
	\label{fig:onset-of-melting}
\end{figure}

To illustrate that our AL approach is applicable to other, non-MD simulation, we use AL to relax (i.e., find the nearest local minimum of) the configurations selected in the training set while learning a $900K$ MD from the previous test.
Relaxation will additionally give us information about the composition of the training set: the configurations learned from MD at $450 K$ should preserve crystalline structure and relax to a perfect crystal, while the configurations learned from MD at $900 K$ should correspond to the liquid phase and relax to disordered (glass-like) configurations.

Thus, we take each configuration from the training set, relax while learning on the fly, and compare its energy to the energy of a relaxed ideal crystalline configurations.
Only 1 of 100 configurations required additional QM calculations while relaxing,
which indicates that there were practically no ``new'' configurations in the process.
The result is shown in Figure~\ref{fig:onset-of-melting}.
We see that 24 of 100 configurations in the training set are perfect crystals, some configurations correspond to the onset of melting (solid-liquid coexistence), and some correspond to fully liquid configurations.

It should be noted that extra care should be taken with regards to the fact that the interatomic potential may slightly change while learning on the fly during relaxation.
Indeed, if a potential increased after treating a certain configuration, then this configuration may be falsely considered as a local minimum (since after the change in the potential energy, the nearby configurations have higher energies).
This issue can be fixed either by re-running the relaxation or formulating the stopping criterion in terms of forces only.

\section{Discussion}\label{sec:discussion}

Our results suggest that the proposed learning-on-the-fly algorithms do not, essentially, reduce the accuracy of interatomic potentials while always keeping the MD trajectory within the physical region.
It should also be emphasized that the AL algorithm introduces a computational overhead that, at least in our test examples, was less expensive than calculating the energy, forces and stresses.
The overhead of retraining our potential was also small compared to the time of calculating the energy, forces and stresses.

Our AL algorithms is not specific to MD and can, in principle, be used with any other type of atomistic simulation, such as structure relaxation, Monte-Carlo sampling, nudge elastic band \cite{1995NEB}, or the accelerated MD methods \cite{Voter1998acceleratedMD}.
Also, our algorithms can learn configurations with varying number of atoms.
This may be important in many applications, including computational structure prediction \cite{Dolgirev2016} where the sought configurations may be of unknown size.

The choice of the D-optimality criterion, which our AL algorithms are based on, was motivated by reducing uncertainty in the parameters of potential.
However, there is also another interpretation.
For example, the elements of the matrix $\mA$ in \QS{1}, $b_j(x^{(k)})$, can be considered as descriptors of the configuration $x^{(k)}$, each configuration is characterized by an $m$-dimensional descriptor vector.
In this sense the D-optimality criterion maximizes the volume of the simplex in $\bbR^m$ formed by $m$ descriptor vectors. In the same way $B_j\big(\br_i^{(k)}\big)$, which are the elements of matrix $\mA$ in \QS{3}, can be considered as the descriptors of a neighborhood of atom $i$ of $k$-th configuration.
Therefore, \QS{3} ``catches'' configurations with the most different atomic neighborhoods in the sense of the D-optimality criterion. This property of \QS{3} can be useful in designing algorithms of a learning-on-the-fly MD with million or more atoms where training has to be done on local environments completed to small configurations.


\section{Conclusion}\label{sec:conclusion}

Machine learning interatomic potentials offer a promising way of combining the accuracy of quantum mechanics and the computational efficiency of the empirical interatomic potentials.
However, the weak point of the machine learning interatomic potentials is reliability---the more general and accurate they are required to be, the harder it is to generate offline the training dataset that ensures no extrapolation at the online evaluation stage.
In the present work we have shown that this problem can be solved by applying active learning to the fitting of the machine learning interatomic potentials.

We have proposed a new active learning scheme based on the D-optimality criterion and have shown empirically that it yields an accurate, computationally efficient, and reliable interatomic interaction model.
In particular, using active learning in the learning-on-the-fly scenario fully resolves the transferability problem---active learning detects when extrapolation is attempted and retrains the potential on those configurations.
In the case when learning on the fly cannot be performed, the proposed active learning techniques allow one to control the degree of extrapolation.
The software, test cases, and examples of usage are published at \url{http://gitlab.skoltech.ru/shapeev/mlip/}.

\section*{Acknowledgments}
	
	The authors thank Prof.\ Ivan Oseledets for useful discussions of application of the D-optimality criterion in active learning, in particular for directing our attention to the maxvol algorithm \cite{oseledets2010how-to1770566}.
	This work was supported by the Skoltech NGP Program No.\ 2016-7/NGP (a Skoltech-MIT joint project).
	A part of the work was done by A.S.\ during the Fall 2016 long program at the Institute of Pure and Applied Mathematics, UCLA.

\section*{References}

\bibliography{paper}

\section*{Appendix A: Functional Form of MTPs}\label{app:mtp}

We used the Moment Tensor Potentials (MTPs) \cite{Shapeev2016-MTP} that have the following functional form
\[
V(\br_i) = \sum_{j=1}^m \theta_j B_j(\br_i),
\]
where $\theta_j$ are the model parameters and $B_j$ are the basis functions indexed by $k\times k$ integer symmetric matrices $\alpha_j\in\bbN^{k\times k}$,  (where $\bbN=\{0,1,\ldots\}$), $j=1,\ldots,m$ as follows.
We let $\calN := \{1,\ldots,N\}$, $\alpha\in\bbN^{k\times k}$ and define, with a slight abuse of notation, the functional form of the basis functions by
\begin{eqnarray}
B_\alpha(\br_i) =
\sum_{\gamma\in\calN^k}
\left(\prod_{\substack{\ell,m=1 \\ m<\ell}}^{k} (r_{i,\gamma_m}\cdot r_{i,\gamma_\ell})^{\alpha_{m,\ell}}
\right) \Bigg(\prod_{\ell} f_{\alpha_{\ell,\ell}}(|r_{i,\gamma_\ell}|)
\Bigg), \nonumber
\end{eqnarray}	
where $f_\mu=f_\mu(\rho)$ ($\mu\in\bbN$) is the set of radial basis functions that vanish for $\rho>R_\cut$.
If, for the sake of argument, we chose $f_\mu(\rho)=\rho^{2\mu}$ then $B_\alpha$ is a polynomial of degree $\sum_{i,j=1}^k \alpha_{i,j}$.
Therefore, the set of $\alpha$ is chosen to include all $\alpha$ such that $\sum_{i,j=1}^k \alpha_{i,j} \leq {\rm deg}$, where ${\rm deg}$ thus limits the ``degree'' of $B_\alpha$.
From the definition it seems that computing $B_\alpha$ is very expensive, however, there is a fast algorithm for computing $B_\alpha$ and their derivatives, refer for the details to \cite{Shapeev2016-MTP}.
In particular, $B_\alpha$ are expressed as tensorial contractions of the following tensor-valued descriptors
\[
M_{\mu,\nu}(\br_i) = \sum_{j} f_{\mu}(r_{ij}) \, \underbrace{r_{ij}\otimes\ldots\otimes r_{ij}}_{\nu\text{ times}},
\]
where $r_{ij}\otimes\ldots\otimes r_{ij}$ is the outer product of $r_{ij}$ with itself $\nu$ times, and $\mu,\nu$ depend on $\alpha$.
These descriptors have an interpretation of moments of inertia of the atomic environment of $i$th atom in the following sense.
If $f_{\mu}(r_{ij})$ is the weight of the atom $j$ in the neighborhood of the atom $i$ then $M_{\mu,0}$ is the mass of the neighborhood (or, in other words, the $0$-th moment of inertia), $M_{\mu,1}$ is the vector of the first moments of inertia (then the center of mass of the neighborhood relative to the atom $i$ is $M_{\mu,1}/M_{\mu,0}$), $M_{\mu,2}$ is the matrix of the second moments of inertia, etc.

\section*{Appendix B: Maxvol algorithm}\label{app:maxvol}

The query strategies \QS{2} and \QS{3} require finding the $m \times m$ submatrix $\mA$ with maximal $|{\rm det}\,\mA|$ (or, in other words, with the maximal volume) in an $k \times m$ matrix $\mB$.
This is done by the maxvol algorithm \cite{oseledets2010how-to1770566}.
This algorithm is based on greedy selection of rows from $\mB$.
Each iteration of this algorithm has $O(m k)$ complexity. The algorithm is as follows. 

\begin{enumerate}  
	\item Start with an initial (e.g., randomly chosen) $m\times m$ submatrix $\mA$ of  $\mB$ and calculate $\mC=\mB\mA^{-1}$.
	\item Find the maximal by absolute value element $\mC_{ij}$ in this matrix.
	\item If $|\mC_{ij}|>\gamma_\th$ then:
	\begin{itemize}
		\item[3.1] swap the $i$-th row of $\mA$ with the $j$-th row of $\mB$,
		\item[3.2] update $\mC = \mB\mA^{-1}$ using the Sherman-Morrison \cite{rank-1-update} rank-one update,
		\item[3.3] go to step 2.
	\end{itemize}
\end{enumerate}

The smaller the threshold parameter $\gamma_\th > 1$ is, the larger $|{\rm det}\,\mA|$ will be, at the cost of making more iterations.

\section*{Appendix C: Active Learning from a Database}\label{sec:learn-from-database}

In this section we show that AL offers some advantages when learning from a given database as opposed to passive learning. 

\subsection*{Errors of fitting}	

We compare the errors of the fitting of the MTP with 100 basis functions using five different strategies, namely fitting on the entire database (passive learning), selecting 100 random configurations, and three AL approaches.
The result for the random selection was averaged over 16 independent samples. 

\begin{table}
	\centering
	\caption{Errors of fitting of different AL approaches.
	}
	\label{tab:table2}
	{\small
	\begin{tabular}{rrcccccc}
		\hline
		Query&& \multicolumn{2}{c}{energy error} & \multicolumn{2}{c}{force error} & \multicolumn{2}{c}{stress error} \\
		strategy &  $\#X_{\rm TS}$ & \multicolumn{2}{c}{(meV/atom)} & \multicolumn{2}{c}{(eV/\AA)} & \multicolumn{2}{c}{(GPa)} \\
		& & RMS & max & RMS & max & RMS & max \\
		\hline
		passive   & 24000 & $0.19$ &$0.82$& 0.016   & 0.13   &  0.052    &  0.13   \\
		random    &   100 & $0.21$ &$0.92$& 0.017   & 0.28   &  0.057    &  0.14   \\ 
		\QS{1}    &   100 & $0.21$ &$0.78$& 0.016   & 0.10   &  0.053    &  0.13  \\ 
		\QS{2}    &    84 & $0.25$ &$0.89$& 0.016   & 0.10   &  0.056    &  0.13  \\
		\QS{3}    &    92 & $0.23$ &$0.79$& 0.016   & 0.09   &  0.057    &  0.13   \\ 
		\hline
	\end{tabular}
	}
\end{table}

As can be seen from Table~\ref{tab:table2}, all AL methods yield marginal increase in the RMS error as compared to passive learning and a significant decrease in the maximal error.
Also, they produce smaller errors than the random selection query strategy.	This indicates the efficiency of the AL methods.
Thus, applying the proposed AL methods to offline learning allow one to reduce the size of $X_{\rm TS}$ (resulting in acceleration of the training stage) while keeping the accuracy essentially at the same level as for the fitting on the entire evaluation set.		

\subsection*{Reliability}	

Finally, we have performed a test of reliability when the potentials are trained offline.
We use the potentials from the previous test fitted on MD trajectories at $T=300$\,K and use them in MD at $T=300$\,K and $T=450$\,K.
We measure the \emph{failure time}, i.e., simulation time until the minimal interatomic distance becomes less then $1.5$\AA.
We performed 100 MD runs and calculated the expected failure time for different MTPs.
These results are presented in Table~\ref{tab:table4}.


\begin{table}[htbp]
	\centering
	\caption{Average failure time, i.e., simulation time until some bond is compressed to $1.5$\AA.}
	\label{tab:table4}
	\begin{tabular}{rrcc}
		\hline
		Query &             &  \multicolumn{2}{c}{failure time (ns)} \\
		strategy & \#$X_{\rm TS}$ &  $T=300$\,K& $T=450$\,K\\
		\hline
		passive	& 24000 & 0.18 & 0.06\\
		random	&   100 & 0.17 & 0.03\\
		\QS{1}	&   100 & 0.66 & 0.04\\
		\QS{2}	&    84 & 1.29 & 0.06\\
		\QS{3}	&    92 & 3.84 & 0.16\\
		\hline
	\end{tabular}
\end{table}

As one can see, the actively learned MTPs are more stable as compared to the passively learned ones. In other words, they have better \emph{transferability}---the ability of the potential to make a prediction for configurations sufficiently different from those in the training set. We argue that this is achieved because AL, in some sense, selects the ``most different'' configurations for training, resulting in a smaller degree of extrapolation if used on a different evaluation set.
Interestingly the neighborhood-based AL, \QS{3}, shows by far the best results then the other two AL approaches.
We speculate that this is because the failure most likely happens locally, when two atoms come too close, and \QS{3} naturally selects such configurations for training.

\end{document}